\documentclass[twoside,twocolumn,9pt]{article}
\pdfoutput=1
\usepackage{extsizes}
\usepackage[super,sort&compress,comma]{natbib} 
\usepackage[version=3]{mhchem}
\usepackage[left=1.5cm, right=1.5cm, top=1.785cm, bottom=2.0cm]{geometry}
\usepackage{balance}
\usepackage{mathptmx}
\usepackage{sectsty}
\usepackage{graphicx} 
\usepackage{lastpage}
\usepackage[format=plain,justification=justified,singlelinecheck=false,font={stretch=1.125,small,sf},labelfont=bf,labelsep=space]{caption}
\usepackage{float}
\usepackage{fancyhdr}
\usepackage{fnpos}
\usepackage[english]{babel}
\usepackage{bm}
\addto{\captionsenglish}{%
  
}
\usepackage{array}
\usepackage{droidsans}
\usepackage{charter}
\usepackage[T1]{fontenc}
\usepackage[usenames,dvipsnames]{xcolor}
\usepackage{setspace}
\usepackage[compact]{titlesec}
\usepackage{hyperref}

\definecolor{cream}{RGB}{222,217,201}

\begin{document}

\pagestyle{fancy}
\thispagestyle{plain}
\fancypagestyle{plain}{
\renewcommand{\headrulewidth}{0pt}
}

\makeFNbottom
\makeatletter
\renewcommand\LARGE{\@setfontsize\LARGE{15pt}{17}}
\renewcommand\Large{\@setfontsize\Large{12pt}{14}}
\renewcommand\large{\@setfontsize\large{10pt}{12}}
\newcommand{\revision}[1]{\textcolor{BrickRed}{{#1}}}
\renewcommand\footnotesize{\@setfontsize\footnotesize{7pt}{10}}
\makeatother

\renewcommand{\thefootnote}{\fnsymbol{footnote}}
\renewcommand\footnoterule{\vspace*{1pt}%
\color{cream}\hrule width 3.5in height 0.4pt \color{black}\vspace*{5pt}} 
\setcounter{secnumdepth}{5}

\makeatletter 
\renewcommand\@biblabel[1]{#1}            
\renewcommand\@makefntext[1]%
{\noindent\makebox[0pt][r]{\@thefnmark\,}#1}
\makeatother 
\renewcommand{\figurename}{\small{Fig.}~}
\sectionfont{\sffamily\Large}
\subsectionfont{\normalsize}
\subsubsectionfont{\bf}
\setstretch{1.125} 
\setlength{\skip\footins}{0.8cm}
\setlength{\footnotesep}{0.25cm}
\setlength{\jot}{10pt}
\titlespacing*{\section}{0pt}{4pt}{4pt}
\titlespacing*{\subsection}{0pt}{15pt}{1pt}

\fancyfoot{}
\fancyfoot[LO,RE]{\vspace{-7.1pt}\includegraphics[height=9pt]{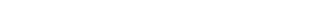}}
\fancyfoot[CO]{\vspace{-7.1pt}\hspace{13.2cm}\includegraphics{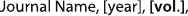}}
\fancyfoot[CE]{\vspace{-7.2pt}\hspace{-14.2cm}\includegraphics{head_foot/RF}}
\fancyfoot[RO]{\footnotesize{\sffamily{1--\pageref{LastPage} ~\textbar  \hspace{2pt}\thepage}}}
\fancyfoot[LE]{\footnotesize{\sffamily{\thepage~\textbar\hspace{3.45cm} 1--\pageref{LastPage}}}}
\fancyhead{}
\renewcommand{\headrulewidth}{0pt} 
\renewcommand{\footrulewidth}{0pt}
\setlength{\arrayrulewidth}{1pt}
\setlength{\columnsep}{6.5mm}
\setlength\bibsep{1pt}

\makeatletter 
\newlength{\figrulesep} 
\setlength{\figrulesep}{0.5\textfloatsep} 

\newcommand{\topfigrule}{\vspace*{-1pt}%
\noindent{\color{cream}\rule[-\figrulesep]{\columnwidth}{1.5pt}} }

\newcommand{\botfigrule}{\vspace*{-2pt}%
\noindent{\color{cream}\rule[\figrulesep]{\columnwidth}{1.5pt}} }

\newcommand{\dblfigrule}{\vspace*{-1pt}%
\noindent{\color{cream}\rule[-\figrulesep]{\textwidth}{1.5pt}} }

\makeatother

\twocolumn[
  \begin{@twocolumnfalse}
{\includegraphics[height=30pt]{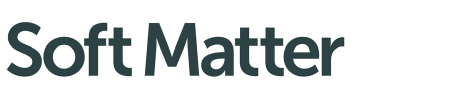}\hfill\raisebox{0pt}[0pt][0pt]{\includegraphics[height=55pt]{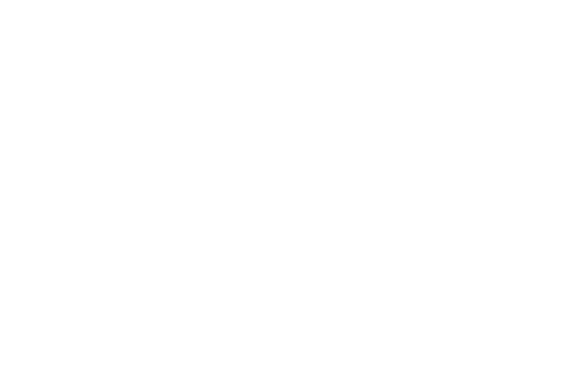}}\\[1ex]
\includegraphics[width=18.5cm]{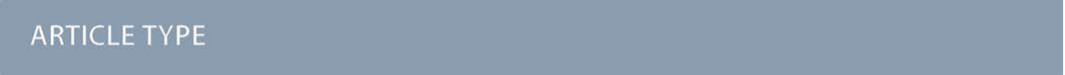}}\par
\vspace{1em}
\sffamily
\begin{tabular}{m{4.5cm} p{13.5cm} }

\includegraphics{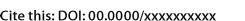} & \noindent\LARGE{\textbf{Memory effects, arches and polar defect ordering at the cross-over from wet to dry active nematics $^\dag$}} \\
\vspace{0.3cm} & \vspace{0.3cm} \\
& \noindent\large{Mehrana R. Nejad,$^{\ast}$\textit{$^{a}$} Amin Doostmohammadi,\textit{$^{b\ddag}$} and Julia M. Yeomans\textit{$^{a}$}} \\

\includegraphics{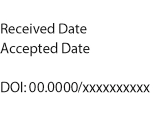} & \noindent\normalsize{We use analytic arguments and numerical solutions of the continuum, active nematohydrodynamic equations to study how friction alters the behaviour of active nematics. Concentrating on the case where there is nematic ordering in the passive limit, we show that, as the friction is increased, memory effects become more prominent and  $+1/2$ topological defects leave increasingly persistent trails in the director field as they pass. The trails are preferential sites for defect formation and they tend to impose polar order on any new $+1/2$ defects. In the absence of noise and for high friction, it becomes very difficult to create defects, but trails formed by any defects present at the beginning of the simulations persist and organise into parallel arch-like patterns in the director field. We show aligned arches of equal width are approximate steady state solutions of the equations of motion which co-exist with the nematic state. We compare our results to other models in the literature, in particular dry systems with no hydrodynamics, where trails, arches and polar defect ordering have also been observed.} \\

\end{tabular}

 \end{@twocolumnfalse} \vspace{0.6cm}
]

\renewcommand*\rmdefault{bch}\normalfont\upshape
\rmfamily
\section*{}
\vspace{-1cm}


\footnotetext{\textit{$^{a}$~The Rudolf Peierls Centre for Theoretical Physics, Department of Physics, University of Oxford, Parks Road, Oxford OX1 3PU, UK.}}
\footnotetext{\textit{$^{b}$~The Niels Bohr Institute, University of Copenhagen, Copenhagen, Denmark. }}
\footnotetext{\dag~Electronic Supplementary Information (ESI) available: [movie captions]. See DOI: 00.0000/00000000.}




\section{Introduction}

Active systems, such as cells or self-propelled colloids, are inherently out of equilibrium  as they continuously take energy from their environment and convert the energy into mechanical stress \cite{julicher2018hydrodynamic,marchetti2013hydrodynamics}. As a consequence of exerting forces on their surroundings, active units can develop new types of nonequilibrium interactions leading to novel self-organisation phenomena \cite{furthauer2019self,ladoux2017mechanobiology,lushi2014fluid,schaller2010polar,ramaswamy2010mechanics,bain2019dynamic}. These are crucial for vital biophysical processes such as organ development~{\cite{ferreira2017physical}}, collective cell migration~{\cite{poujade2007collective}, subcellular organisation~\cite{needleman2014determining}}, and bacterial biofilm formation~{\cite{dell2018growing}}. Moreover the active self-organisation has proven instrumental in inspiring the design of new active synthetic materials, capable of self-healing and self-propulsion~\cite{khadka2018active,lozano2016phototaxis}, and,
 from the fundamental physics standpoint, active systems provide a testing ground for building and examining theories of out-of-equilibrium statistical physics \cite{solon2015pressure,fodor2016far,pietzonka2019autonomous,PhysRevLett.114.198301}. These widespread implications of active systems in various disciplines and across several length scales call for generic frameworks for studying and understanding active materials.
 
Research to understand the properties of active materials has suggested that it is often useful to divide them into two major categories, dry active matter and wet active matter.  Dry models are defined as those in which momentum is not conserved, usually due to the friction with a substrate \cite{marchetti2013hydrodynamics}.
Active systems that have been described by dry models include shaken granular matter \cite{junot2017active,kumar2014flocking} and self-propelled robots \cite{deblais2018boundaries} where interactions are mediated through collision events instead of fluid flows. A system of bacteria moving on a frictional substrate has also been considered using a dry active matter model ~\cite{meacock2020bacteria}. 

By contrast, in wet models momentum is conserved. In wet active systems active particles, such as microswimmers \cite{blaschke2016phase} are surrounded by a fluid medium, usually incompressible water, and as a result the forces between individuals are mediated through the fluid.
 Wet active systems include microtubules driven by kinesin motors \cite{hardouin2019reconfigurable,sanchez2012spontaneous}, actomyosin suspensions \cite{rupprecht2018maximal,seara2018entropy}, swimming micro-organisms \cite{blaschke2016phase} and artificial swimmers \cite{brosseau2019relating,huang2019adaptive,lozano2016phototaxis}. Of particular importance, the presence of the fluid mediates long-range hydrodynamic interactions between active particles in wet systems, which do not exist in their dry counterparts \cite{simha2002hydrodynamic,PhysRevLett.100.178103,Wensink14308}.   
Theoretically, wet and dry active matter are often studied in isolation: hydrodynamic theories, which include long-range interactions, are successful in describing wet systems, but often fail to accurately capture the properties of dry active matter; while completely dry treatments neglect hydrodynamic effects altogether and are not capable of representing wet systems. It is, therefore, interesting to search for frameworks that allow for a unified representation of wet and dry active systems, and to identify the limits of such unifying approaches.

Since the main factor distinguishing between wet and dry systems is the presence or absence of momentum conservation, introducing hydrodynamic screening to wet systems could provide a link to their dry counterparts. 
 Possible ways to do this are by increasing the friction with an underlying substrate or by changing the viscosity of a neighbouring fluid which provides a momentum sink \cite{PhysRevE.97.063101}. Indeed it is  possible to impose anisotropic friction by interfacing active suspensions with smectic liquid crystals \cite{guillamat2016control}.
In general, friction screens the flow field created by active suspensions thus introducing a length-scale over which the velocity field goes to zero. 
This raises the questions of whether increasing friction can engender a crossover between wet and dry active systems and the extent to which the high friction limit matches the behaviour of dry active matter. 

To address these questions, in this paper we consider the role of friction in active nematohydrodynamics, the canonical theory of wet active materials. This model does not conserve momentum in the presence of friction, however, the frictionless equations do obey conservation laws. We study the behaviour of the director field and $\pm 1/2$ defect with increasing friction and show that when friction is increased in an incompressible active suspension in the nematic phase, memory effects develop. As $+1/2$ motile topological defects move through the system they leave long-lived distortions in the director field which can affect the motion of other defects, leading to the polar defect ordering that has been observed in microscopic simulations \cite{patelli2019understanding,decamp2015orientational}.
At very high friction all motile defects annihilate after some time to leave arches, periodic modulations of the nematic director with polar symmetry. We show analytically that arch patterns are steady state solutions of the equations, which coexist with nematic ordering in agreement with the completely dry simulations in \cite{patelli2019understanding}. Our results also demonstrate that the polar ordering of defects and polar modulations in the nematic director are general behaviours of active nematic materials at high friction, and are not restricted to the compressible systems studied in \cite{putzig2016instabilities,oza2016antipolar,decamp2015orientational,shankar2019hydrodynamics,srivastava2016negative}.

Section~\ref{activen} of the paper introduces the active nematohydrodynamic equations of motion and summarises the physical behaviour of active nematics, in particular active turbulence and motile topological defects. In Section~\ref{lowT} we describe the dynamics of the nematic phase as friction is increased, concentrating in particular on trails, arch-like structures in the director field formed by the passage of topological defects.  We show that our numerical results agree with a linear stability analysis presented in 
Section~\ref{lsa}, and give analytic arguments demonstrating the metastable nature of the arches in Section~\ref{metastability}. There are several interesting papers in the literature that have identified arches and polar defect ordering in different contexts \cite{putzig2016instabilities,oza2016antipolar,decamp2015orientational,shankar2019hydrodynamics,srivastava2016negative,patelli2019understanding} and in Section~\ref{comparison} we summarise and compare these results. Finally Section~\ref{discussion} concludes the paper, listing the main results and comparing to the effects of friction when the active nematic is in the isotropic phase.

\section{Active nematics}\label{activen}

Many active suspensions, such as swimming bacteria and microtubule-motor suspensions, are in the low Reynolds number regime. As a result, a single active particle can only generate a force dipole in
a momentum conserving system, and hence, in general, generates a nematic far flow field \cite{PhysRevLett.105.168101}. This means that the active stress is described to leading order by a nematic tensor \cite{simha2002hydrodynamic}. Moreover, nematic interactions can also be caused by excluded volume effects between active particles with elongated shape.\\ Therefore, the fundamental continuum equations that describe wet active nematics are the active nematohydrodynamic equations which are coupled equations for the evolution of the nematic tensor, $\textbf{Q}=2 S(\textbf{n}\textbf{n}-\textbf{I}/2)$ in two dimensions, and the associated incompressible fluid velocity, $\mathbf{u}$. These read
\begin{align}
   &\partial_t \textbf{Q} + \textbf{u} \cdot \boldsymbol{\nabla} \textbf{Q}- \boldsymbol{\mathcal{W}}= \gamma \: \textbf{H},\label{n:qevoln}\\
    &\rho \left(\partial_t + \textbf{u}\cdot \boldsymbol{\nabla}\right) \textbf{u}= \boldsymbol{\nabla} \cdot \boldsymbol{\Pi}-\Gamma  \textbf{u} , \quad \boldsymbol{\nabla}\cdot \textbf{u}=0.\label{eqn:ns}
\end{align}
In the definition of the nematic tensor the director field $\textbf{n}$ represents the orientation of the nematic alignment and the magnitude of the nematic order is denoted by $S$. In the evolution of the $\textbf{Q}$ tensor, $\gamma$ is the rotational diffusivity and the molecular field, $\mathbf{H}= K \nabla^2 \mathbf{Q} +A[B-2 Tr(\textbf{Q} \cdot \textbf{Q}) ]\textbf{Q}$, drives the system towards the minimum of a free energy. The first term in the molecular field derives from the energy cost due to distortions in the nematic field, assuming a single Frank elastic constant $K$, and the second term (with $A>0$) accounts for the relaxation of the nematic order parameter to the homogeneous nematic phase with $S=S_0=B^{1/2}$ for $B>0$. The generalized advection term  
 \begin{align}\label{elastic}
    \boldsymbol{\mathcal{W}}=&(\lambda \textbf{E}+\boldsymbol{\Omega})\cdot(\textbf{Q}+\frac{\textbf{I}}{2})+(\textbf{Q}+\frac{\textbf{I}}{2})\cdot(\lambda \textbf{E}-\boldsymbol{\Omega}) -\lambda(\textbf{Q}+\textbf{I})Tr(\textbf{Q} \cdot \textbf{E})
\end{align} 
 models the response of the nematic field to the strain rate $ \textbf{E}$ and vorticity $\boldsymbol{\Omega}$, and $\lambda$ denotes the flow-aligning parameter. 
  
In the Navier-Stokes equation (\ref{eqn:ns}), $\rho$ is the density of the suspension, $\Gamma$ is a friction coefficient, and the stress tensor, $\boldsymbol{\Pi}$, includes viscous, elastic and active contributions. The viscous stress, $\boldsymbol{\Pi}^{v}=2\eta \textbf{E}$, where $\eta$ is the viscosity, and the elastic stress,
\begin{align}\label{ela}
\Pi^{p}_{ij} =& -P \delta_{ij} + \lambda (Q_{ij}+\delta_{ij} ) Q_{kl} H_{kl} -\lambda  H_{ik} (Q_{kj}+\frac{\delta_{kj}}{2})\nonumber \\ & -\lambda(Q_{ik}+\frac{\delta_{ik}}{2}) H_{kj}+ Q_{ik} H_{kj}- H_{ik} Q_{kj} - K (\partial_i Q_{kl})(\partial_j Q_{kl}),
\end{align} 
where  $P$ is the pressure, are familiar terms that appear in the dynamical equations of passive liquid crystals. 
Coarse-graining the dipolar force fields of the active nematogens leads to an active contribution to the stress that characterises wet active nematics, $\boldsymbol{\Pi}^{\text{a}}=\zeta \textbf{Q}$. 
 The active stress  is not isotropic: in extensile systems ($\zeta<0$), it acts to extend a nematic region along its director whereas in contractile materials ($\zeta>0$) it contracts a nematic region along the director. 
 
\begin{figure*}
\centering
  \includegraphics[width=0.85\linewidth]{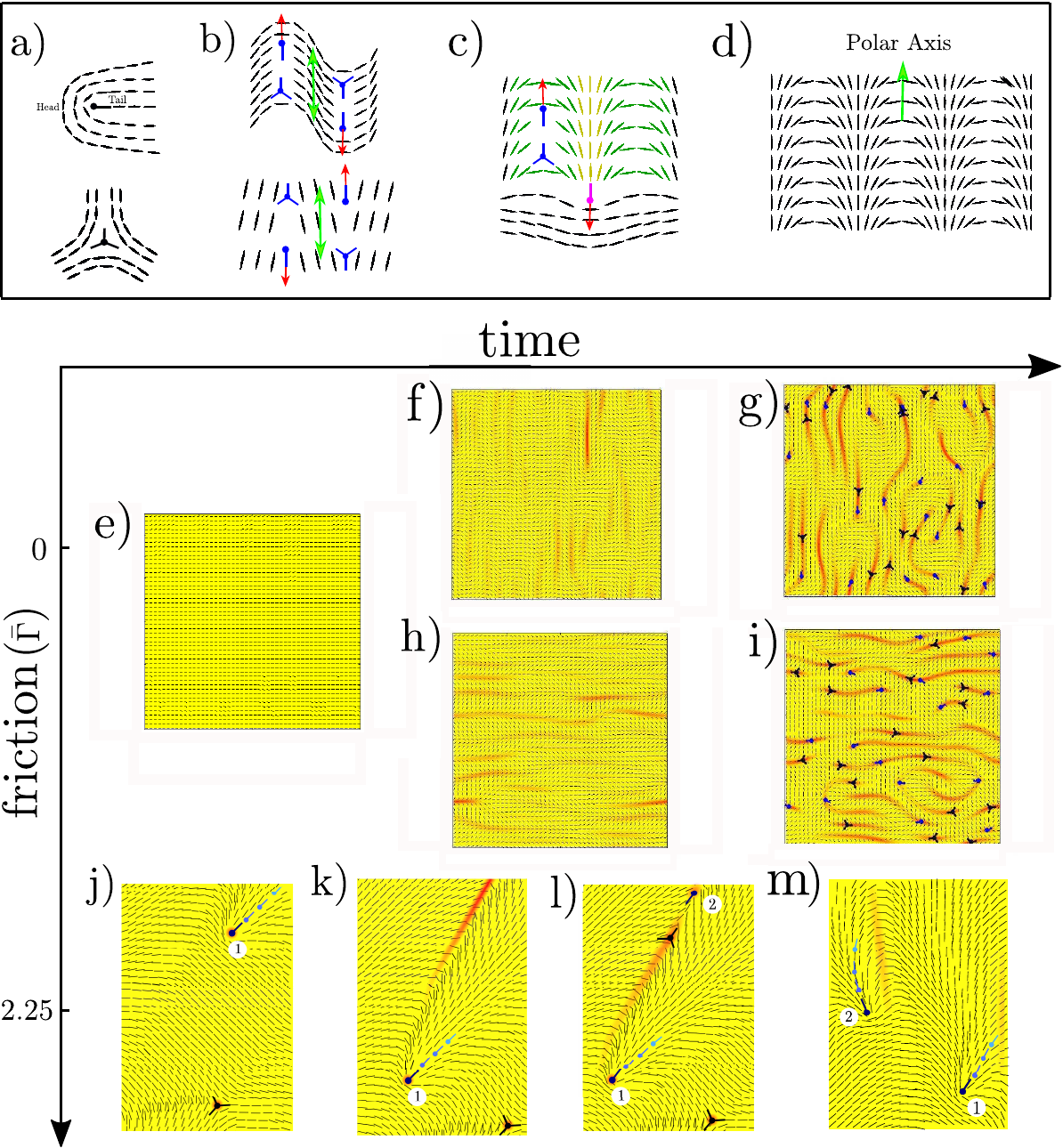}
  \caption{{\bf The mechanism of defect creation in the nematic regime changes with increasing friction:}
  (a) Director field of topological defects in 2D active nematics. Top and bottom represent a $+1/2$ and a $-1/2$ defect, respectively. $+1/2$ defects have a polar symmetry while $-1/2$ defects have three-fold symmetry. (b)-(d) Difference between walls, trails and arches: The direction along walls and arches is represented by green arrows. 
  The difference between walls and arches is that neighbouring arches point in the same direction (Fig. 1d), whereas neighbouring walls point in opposite directions (Fig. 1b). We use the term trails to denote distortions left behind moving defects (Fig. 1c). These are of a similar structure to arches and, in the large friction regime, all defects annihilate and trails form static arches.
  (b) In the absence of friction, pairs of $\pm 1/2$ defects form at walls, which are lines of high distortion in the nematic order. Top (bottom) shows the director field at walls in extensile (contractile) systems. The orientation of pairs of newly created defects is shown in blue and the direction of motion of the $+1/2$ defects by the red arrow. (c) The trail left by an extensile $+1/2$ defect (shown in purple) at intermediate and high friction. The trail incorporates bend and splay distortions represented by green and yellow directors, respectively. At intermediate friction, new pairs of defects (shown in blue) form on the bend region of trails. The red arrow shows the direction of the motion of the defect and green two-headed arrows show the direction of walls. (d) The director field in the presence of three arches, structures that can be formed by the passage of defects at high friction.  The green arrow indicates the polar axis of the arches.
  (e) Direction of nematic order at $t=0$. (f) In an extensile system, walls form perpendicular to the direction of nematic order. (g) As time passes, defect pairs form on the walls, and the $+1/2$ defect moves away from the $-1/2$ defect along the wall. (h) In a contractile system, walls form parallel to the order. (i) Defect pairs also form on the walls for a contractile system.
  (j) - (m) time evolution of a defect creation event at intermediate friction. The $+1/2$ defects specified with $1$ and $2$ are the original and the newly created defects, respectively. The frame is moved with the defects to track both defects and distortions over time. The trajectory of the $+1/2$ defects is represented by the small light blue defects. (j) and (k) show that the system develops a memory and defects trails persist for some time. (l) Shows a creation event. Since distortions are larger along the trails, further creation events tend to occur close to them with $+1/2$ defects forming anti-parallel to the direction of the original $+1/2$ defects. (m) Represents that the new $+1/2$ defect aligns with the original defect after a while.
}
  \label{walls}
\end{figure*}

As a result of the stress that active particles continuously exert on their surroundings, the nematic phase is unstable and evolves towards a  state characterized by chaotic flows, with prominent  vorticity and fluid jets, known as active turbulence \cite{fraden2019turbulent,PhysRevLett.110.228102,shankar2018defect,cortese2018pair,schaller2013topological,thampi2016active,Wensink14308,slomka2017spontaneous,martinez2019selection,PhysRevLett.120.208101,lemma2019statistical}. 
 Experimental and numerical measurements of the vorticity correlations in active turbulence have shown that the vortices are associated with an active length scale, $\sqrt{K/\zeta}$, and that the areas of the vortices are exponentially distributed at large scales ~\cite{giomi2015geometry,guillamat2017taming}. 
Active turbulence is also distinguished by the presence of motile topological defects, singularities in the director field \cite{giomi2014defect,doostmohammadi2016stabilization,decamp2015orientational}. 
In 2D active nematics, topological defects of half-integer charge ($\pm 1/2$), which have the lowest energy~\cite{PhysRevLett.110.228101,shi2013topological}, have been observed in a wide range of experimental systems~\cite{sanchez2012spontaneous,keber2014topology,decamp2015orientational}.
As shown in Fig.~\ref{walls}(a) top, $+1/2$ defects have polar symmetry which means that they can move even in the absence of a background flow when activity breaks time-reversal invariance. Their direction of motion depends on the details of the stress that they exert on their surroundings and can be towards the head (bend region) or the tail (splay region) of the defect~\cite{giomi2014defect}, although in the continuum models considered so far, $+1/2$ defects have always been found to move towards the head in extensile systems and towards the tail in contractile systems. $-1/2$ defects, by contrast, have three-fold symmetry (Fig.~\ref{walls}(a), bottom). They produce flows, but since their director field is not polar, the flows balance at the center of defects and do not lead to self-propulsion. Just as in passive nematics $+1/2$ and $-1/2$ pairs tend to annihilate, but in an active system there is also energy available to create topological defect pairs which then move apart. 
~\\

\section{Trails and arches}\label{lowT}
We consider a wet active system which, in the absence of activity, is in the  nematic phase and solve the equations of motion~(\ref{n:qevoln}--\ref{ela})
numerically to understand how the behaviour of the topological defects changes with increasing friction. The lattice spacing,
time-step and viscosity are taken as unity, and $K = 0.015$, $\gamma=0.4$, $\zeta=\pm0.02$ , $\rho = 40$, $B=1$, $A=0.01$ and $\lambda = 0$, unless otherwise stated. These values of the parameters lead to a Reynolds number of the order of  $Re=\rho u \ell/\eta \sim 10^{-5}-10^{-4}$ in our system.
We define dimensionless time  as $\bar{t}= t \zeta /\eta$ where the active time scale, $\tau_{\text{act}}=\eta/\zeta$, characterizes the  active energy injection. As dimensionless friction we use $\bar{\Gamma}=\Gamma K/(\zeta \eta)$, the 
square of the ratio of the activity-induced length scale $\sqrt{K/\zeta}$ to the frictional screening length $\sqrt{\eta/\Gamma}$. 
We change the value of the dimensionless friction coefficient in the interval $0 \leq \bar{\Gamma} \leq 3$. When we refer to low and intermediate friction, we use $\bar{\Gamma}=0.007$ and $\bar{\Gamma}=2.25$, respectively. For high friction, we use $\bar{\Gamma}=3$, where all the defects in the system annihilate and the system evolves towards the arch state. \\

For low friction, as a result of the active stress, the
nematic phase is unstable to bend (splay) perturbations for extensile (contractile) systems. As the instability grows, disturbances in the director field tend to localise into {\it walls}, lines of high distortion in the director field. The configuration of the nematic director field at the walls is shown in Fig.~\ref{walls}(b), where the green double-headed line indicates the direction of the walls. In an extensile system, walls comprise bend distortions (Fig.~\ref{walls}(b), top) and in contractile systems, they are formed by splay distortions (Fig.~\ref{walls}(b), bottom).  Figs.~\ref{walls}(f) and (h) present snapshots showing that we recover the formation of walls in numerical solutions of Eqs.~(\ref{n:qevoln}--\ref{ela}) in the absence of friction for extensile and contractile systems, respectively. \\

For low values of the friction, the walls provide preferential sites for the formation of $\pm 1/2$ defects. Hence, as shown in Figs.~\ref{walls}(g) and (i), and also in Movies $1$ and $2$, defects form with their polar axes perpendicular to the direction of the initial nematic order in extensile systems and parallel to the direction of the order in contractile systems. This is true at all times, but less obvious as active turbulence becomes well established and the nematic order is more broken up. The defects tend to move along the walls, restoring local nematic order and leading to new rounds of instability.\\
 
Interestingly, we observed that as the friction is increased, the system develops a memory and, rather than restoring nematic order, the motion of $+1/2$ defects leaves persistent trails in the director field. 
As represented in Fig.~\ref{walls}(c), for an extensile $+1/2$ defect moving in the direction of the red arrow, the trail comprises  a splay distortion along the trajectory of the defect (shown in yellow) with two neighbouring regions of bend distortion (shown in green) parallel to the  trajectory. 
At lower values of the friction, $\bar{\Gamma}=1.87$,
the trails stay in the system just for a short while since the passage of defects that move in different directions destroys them (see Movie $3$). As the friction is increased to intermediate values, the distortions left by the defects extend in space and persist for longer times.\\

\begin{figure*}
\centering
  \includegraphics[width=0.8\linewidth]{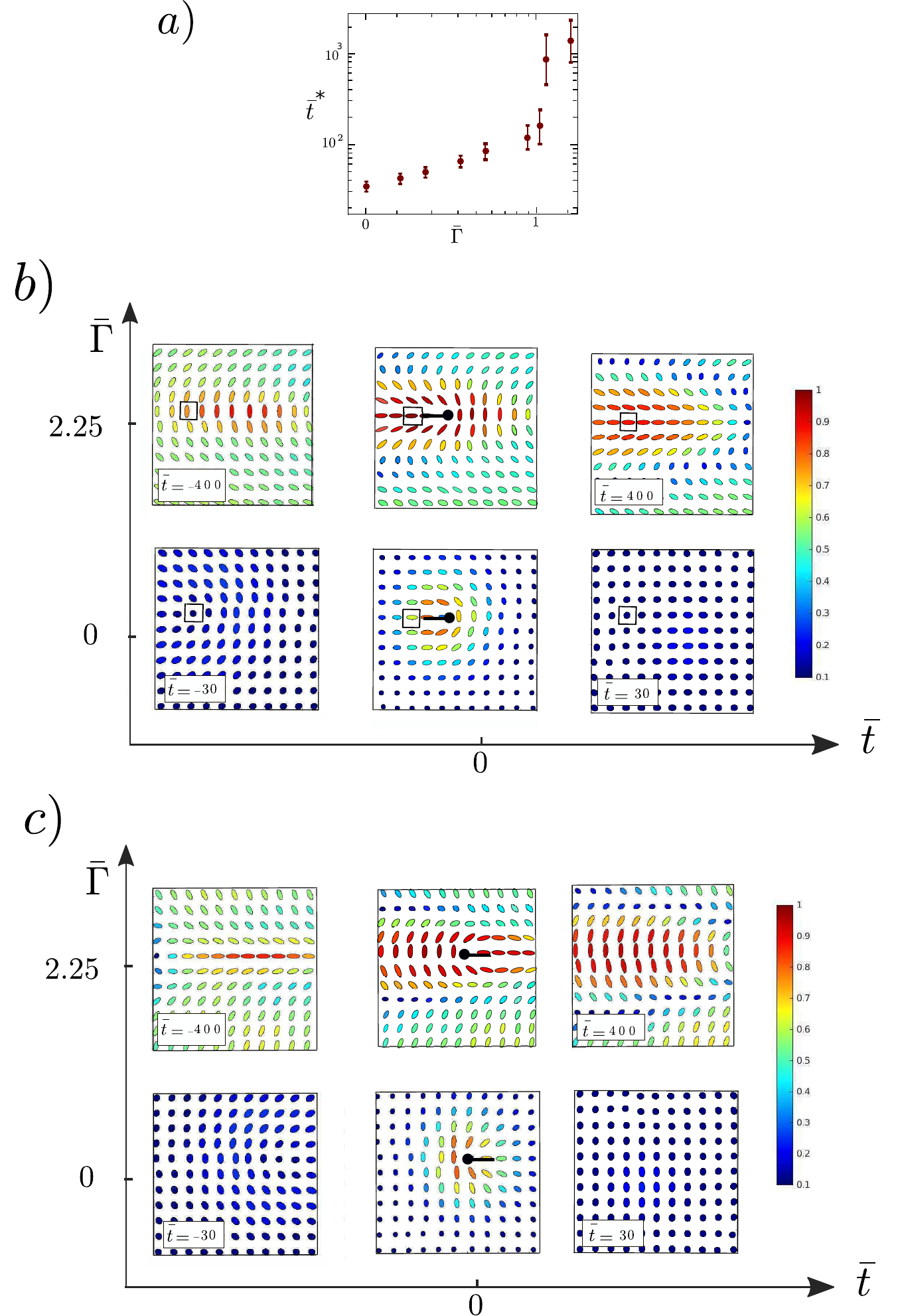}
  \caption{{\bf Effect of friction on the trails left by defects:} (a) Average time over which the nematic order parameter behind defects, in the position specified by the black square, decays to $70\%$ of its initial value, ($S(\bar{t}^*)=0.7 S(\bar{t}=0)$), as a function of friction. The average is taken over many defects.
 Average director around a location where defects pass at $\bar{t}=0$ for (b) extensile and (c) contractile suspensions comparing friction (top, $\bar{\Gamma}=2.25$) to no friction (bottom). Each ellipse represents the director averaged over a $2 \times 2$ square of lattice sites. The defects are at the center of the box at $\bar{t}=0$. Negative times show the average director before a defect enters the box and positive times show times after the defect has left the box. 
}
  \label{lifetime}
\end{figure*}

In order to better characterize the trail formation in Fig.~\ref{lifetime}, we quantify the memory effects. We define $\bar{t}=0$ as the time at which a defect is located at a given position $(x(0),y(0))$, where the $x$-axis is chosen to lie along the trajectory of the defect. We measure the variation of the average nematic order parameter at ($x(0)-d,y(0))$ with $d=5$ lattice units, as a function of time, and define $\bar{t}^*$ as the average time for this quantity to decay to 0.7 of its value at $\bar{t}=0$. \\

The position $(x(0)-d,y(0))$ is shown in Fig. \ref{lifetime} with a black square. The average order parameter, defined in the lab frame, is 
\begin{align}\label{order}
\langle Q_{ij}(\textbf{r},t) \rangle =& \sum_{m=1}^N \frac{Q_{ij,m}(\textbf{r},t)}{N},\nonumber \\  S^2(\textbf{r},t)=&\frac{\langle Q_{xx}(\textbf{r},t) \rangle ^2+\langle Q_{xy}(\textbf{r},t) \rangle ^2}{4},
\end{align}
where the average is taken over $N$ different $+1/2$ defects in a reference frame (chosen independently for each defect) in which the defect is at the centre of the simulation box at $t= 0$ pointing towards the right. At $t=0$ the black square is on the trail of  the defect. As time passes, the defect moves further away from this position towards the right, but its trail stays in the system for a while.\\ 

The results indicate that the trails persist $\sim 100$ times longer for $\bar{\Gamma} > \sim 1$ than for zero friction. Without friction, the nematic order behind defects goes to zero quickly as the ordered state is unstable to small perturbations. By increasing friction, the active flows are damped by friction force and the growth rate of perturbation decreases. The sharp increase of the defect memory time close to $\bar{\Gamma}=1$ corresponds to the point where the screening length scale becomes comparable to the intrinsic activity-induced length scale. Fig.~\ref{lifetime}(b) shows that without friction and in active turbulence, the {\it average} nematic order goes to zero at the defect core and far from the defect. The small value of the nematic order at large distances from the defect is due to the presence of other defects and active turbulence that makes the director uncorrelated between different sample defects at large distances.\\

\begin{figure}
\centering
  \includegraphics[width=0.8\linewidth]{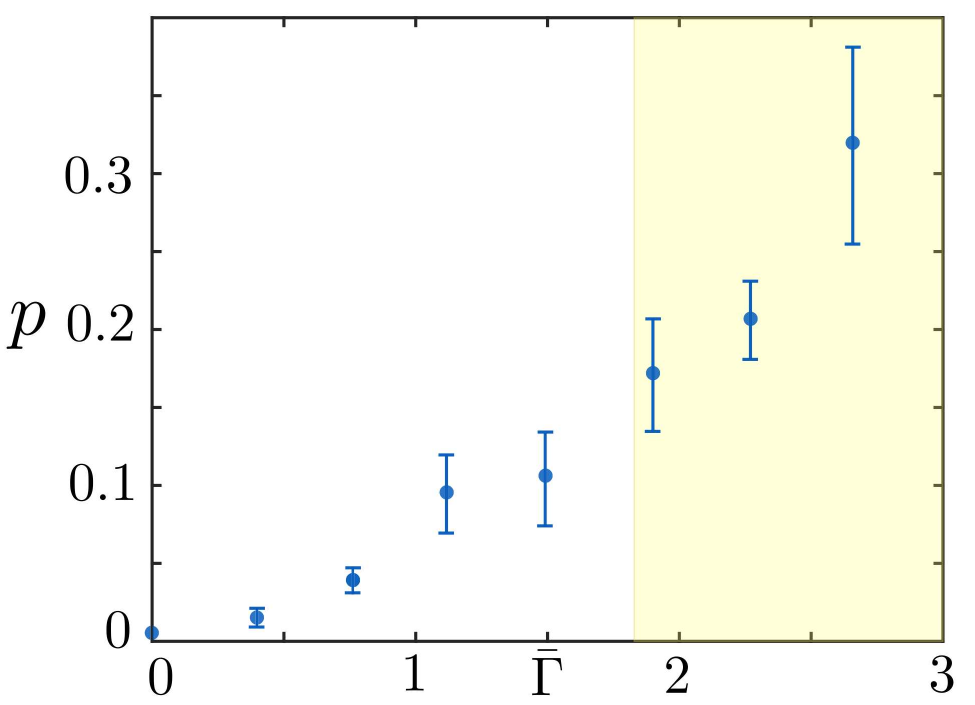}
  \caption{{\bf Effect of friction on the polarisation of $+1/2$ defects:} Increasing friction in the nematic phase increases the polarisation of $+1/2$ defects. In the yellow region, we observe creation events on distortions left by other defects. }
  \label{polar}
\end{figure}

Importantly, once trails are established they, rather than walls, provide the preferential sites for defect formation.  Examples of creation events in the intermediate friction regime are shown in Fig.~\ref{walls}(j)-(m) and Movie $4$.  In the initial stage of its formation, the direction of a new $+1/2$ defect (defect number $2$ in Fig. \ref{walls}(j)) is anti-parallel to the direction of the original $+1/2$ defect that created the trail (defect number $1$), but after some time it rotates and becomes parallel to the original defect, thus leading to polar ordering of $+1/2$ defects (see Movie $4$ and Fig.~\ref{walls} (m)). In Fig. \ref{polar}, we represent the defect polarisation by increasing friction and we mark the regime in which defects form in the wake of existing defects by yellow. \\

Aligning torques due to active flows are primarily responsible for the polar alignment of defects. To demonstrate this we have performed simulations with and without activity starting from an initial condition which includes arches and polar order of $+1/2$ defects (Movies $5$ and $6$, respectively). We then observe the dynamics of a newly created defect which points anti-parallel to the direction of the polarisation of other defects. This defect is marked by a red circle in the movies. In an active system (Movie $5$), the defect becomes parallel to other defects after a while, whereas in a passive system it does not change its direction and annihilates (Movie $6$), showing that polar ordering of the $+1/2$ defects can not be established without the presence of activity. \\

While contractile $+1/2$ defects orient their head toward the polar axis of arches (see Fig.~\ref{walls}(d)), extensile defects point in the opposite direction. As a result, at intermediate frictions, both extensile and contractile defects move in the same direction, antiparallel to the polar axis of the arches. \\

As the defects  move along the trails they re-form them. Extensile (contractile) defects move along bend (splay) regions of the trails and leave splay (bend) distortions in their wake. This is illustrated, and compared to the no friction case, in  Figs.~\ref{lifetime}(b) and (c), where we measure the average director field as a defect moves across the simulation box from left to right. The three frames show: $\bar{t }< 0$ a time  before the defect enters the simulation box,  $\bar{t}=0$ where the defect is at the centre of the box, and $\bar{t} > 0$ a time when the defect has left the box. For zero friction the defects appear at walls and leave nematic order in their wake. For intermediate friction defects form at  trails (see creation events marked by red circles in Movie $4$) and move the arch patterns perpendicular to their direction of motion as they pass (Movies $7$ and $8$). \\

 At yet higher frictions there is not enough energy available to create defect pairs. However, any defects already in the system, for example due to noisy initial conditions, can create arches before annihilating. When all defects annihilate and arches form, the widths of arches equalise (see Movie $9$). The number of arches is conserved during this process and as such there is no preferred width as pointed out in \cite{patelli2019understanding}. We will show in Section~\ref{metastability} that arches with equal width are a steady state solution of the nematohydrodynamic equations with friction. The arch states persist as metastable states which co-exist with the nematic ground state. This is illustrated in Fig.~\ref{archnew2} which compares the stability diagrams of the nematic phase as a function of elastic constant and friction for simulations with small and large noise as initial condition. \\
 
\begin{figure*}
\centering
  \includegraphics[width=0.9\linewidth]{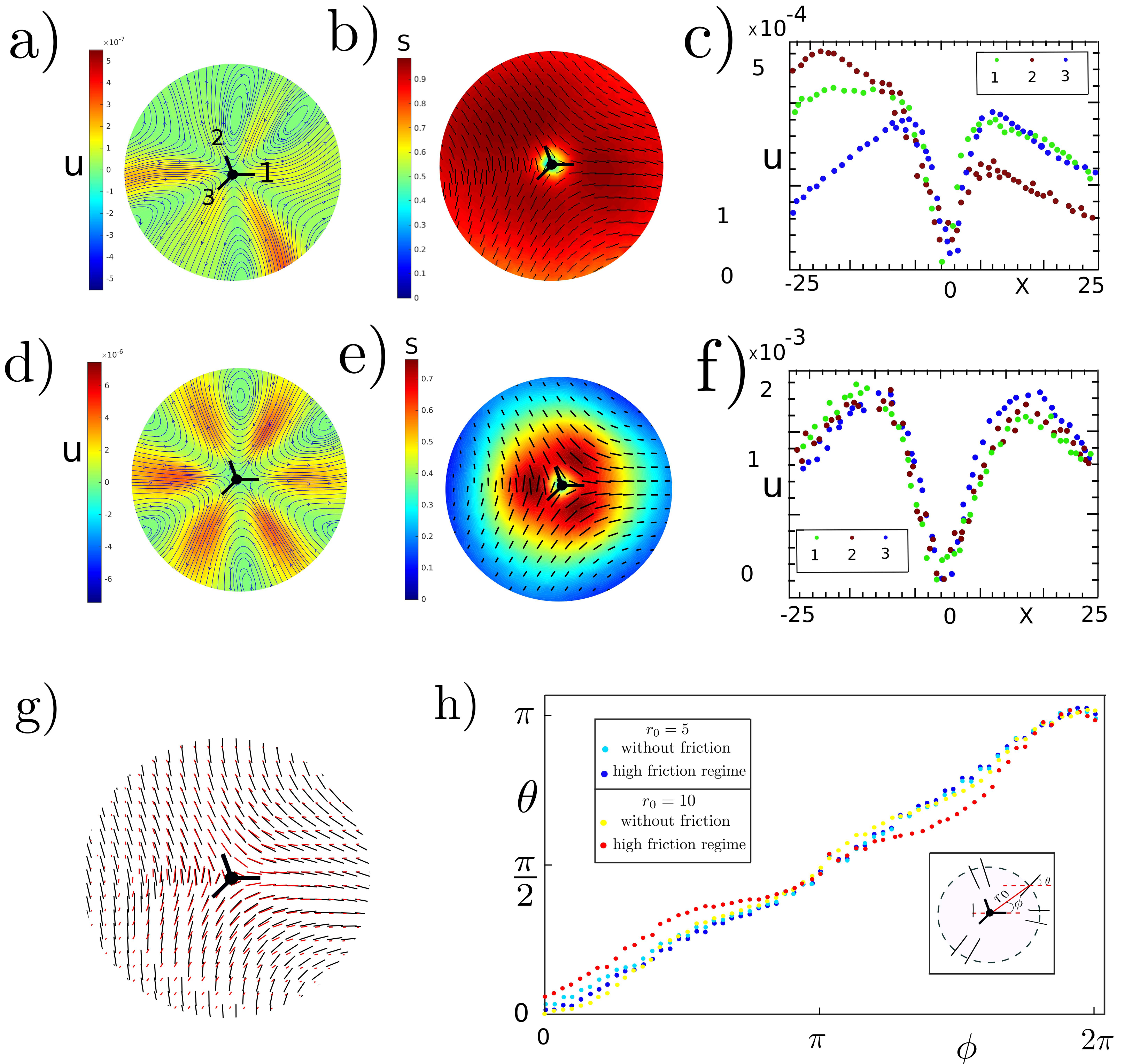}
  \caption{{\bf \bm{$-1/2$} defects move along arches:} (a) average flow field and (b) director field around $-1/2$ defects in the high friction regime. The presence of arches in the high friction regime breaks the three-fold symmetry of $-1/2$ defects at large distances and the active flow moves them along the arches. (c) Average velocity along the three arms of the defects. (d) average flow field and (e) director field around $-1/2$ defects in low friction regime and (f)  the average flow field along the arms, for comparison. (g) Comparison of the average director field around $-1/2$ defects in the high and low friction regimes, represented by black and red respectively. The length of the lines represents the magnitude of the nematic order. (h) Average director field on a circle with radius $r_0$ around the defect. (g) and (h) show that the presence of arches breaks the symmetry of $-1/2$ defects at large distances. All the spatial scales are in lattice Boltzmann units.}
  \label{minushalf}
\end{figure*}

  The flow field and director field around $-1/2$ defects in the high and low friction regimes is represented in Fig.~\ref{minushalf}. As shown by this figure, in the high friction regime, the three fold symmetry of the flows around $-1/2$ defects is broken by the arch structures at large distances and, as a result, the active flow moves the $-1/2$ defects anti-parallel to the arches (See Movie $7$). In Fig.~\ref{minushalf}(g) we compare the average director field around $-1/2$ defects in the low and high friction regimes (shown in red and black, respectively). This shows that arches break the symmetry of the $-1/2$ defects at large distances but do not change the average director field close to the centre of the defects. To show this more clearly, in Fig.~\ref{minushalf}(h), we plot the average direction of the director field on a circle with radius $r_0$ around the $-1/2$ . In active turbulence, the shape of $-1/2$ defects does not change by increasing the radius from $r_0=5$ to $r_0=10$. However, in the large friction regime, the shape of defects is similar to active turbulence at small distances ($r_0=10$) but deviates from it at larger distances ($r_0=10$). \\
 
 We next show that this data is in agreement with a linear stability analysis, and then demonstrate the coexistence of arch and nematic steady states. 
 
\section{Linear stability analysis}\label{lsa}
We now consider an incompressible suspension in the presence of friction and study the stability of the homogeneous nematic phase. We assume that the director is oriented along the $x$-direction and follow the evolution of small perturbations over time. 
The elements of the perturbed nematic tensor are $Q_{xx}=Q_{xx}^0+\delta Q_{xx}$ and $Q_{xy}= Q_{xy}^0+\delta  Q_{xy}$ where $(Q^0_{xx},Q_{xy}^0)=(S_0/2,0)$. Using Eqs. (\ref{n:qevoln}) and (\ref{eqn:ns}), and representing the Fourier transform of any fluctuating field $\delta f$ as $\delta f(\textbf{r},t)=\int d \textbf{q}\: \tilde{f}(\textbf{q},t)\: e^{i \textbf{q} \cdot \textbf{r}}$, the evolution of the perturbations in the low-Reynolds number limit are
\begin{align}
     \partial_t \tilde{Q}_{xx}&= \tilde{Q}_{xx}\left\{(-K q^2-2 S_0^2 A)\gamma +(\Gamma +\eta q^2)^{-1} M\right\}\nonumber\\&+ \tilde{Q}_{xy} \left\{\zeta \lambda (1-{S_0^2}/{2}) q^2 \sin 4 \theta (4\Gamma +4\eta q^2)^{-1}\right\},\\
     M &=\lambda q^2(1-{S_0^2}/{2})\cos^2\theta (\cos 2\theta-1)[\zeta+2 S_0^2\lambda A (1-{S_0^2}/{2})],\nonumber\\
     \partial_t \tilde{Q}_{xy}&= (\Gamma +\eta q^2)^{-1}\tilde{Q}_{xx} q_x q_y (S_0 +\lambda \cos 2\theta)\left\{\zeta + 2 S_0^2 \lambda A (1-{S_0^2}/{2})\right\}\nonumber\\& -q^2 \tilde{Q}_{xy}\left\{K \gamma+(S_0 +\lambda \cos 2\theta)\zeta \cos2\theta (2\Gamma +2\eta q^2)^{-1} \right\}.
\end{align}

For $\lambda=0$, these equations simplify to
\begin{eqnarray}\label{eq1a}
\partial_t \tilde{Q}_{xx}&=& - \tilde{Q}_{xx}(K q^2+2 S_0^2 A)\gamma,\label{sa}\\ \partial_t \tilde{Q}_{xy}&=&-q^2 \tilde{Q}_{xy}\left\{K \gamma+S_0\zeta \cos 2\theta (2\Gamma +2\eta q^2)^{-1} \right\}\nonumber \\&&+(2\Gamma +2\eta q^2)^{-1} \tilde{Q}_{xx} q^2 \sin 2\theta S_0 \zeta.\label{sa2}
\end{eqnarray}

\begin{figure*}
\centering
  \includegraphics[width=0.9\linewidth]{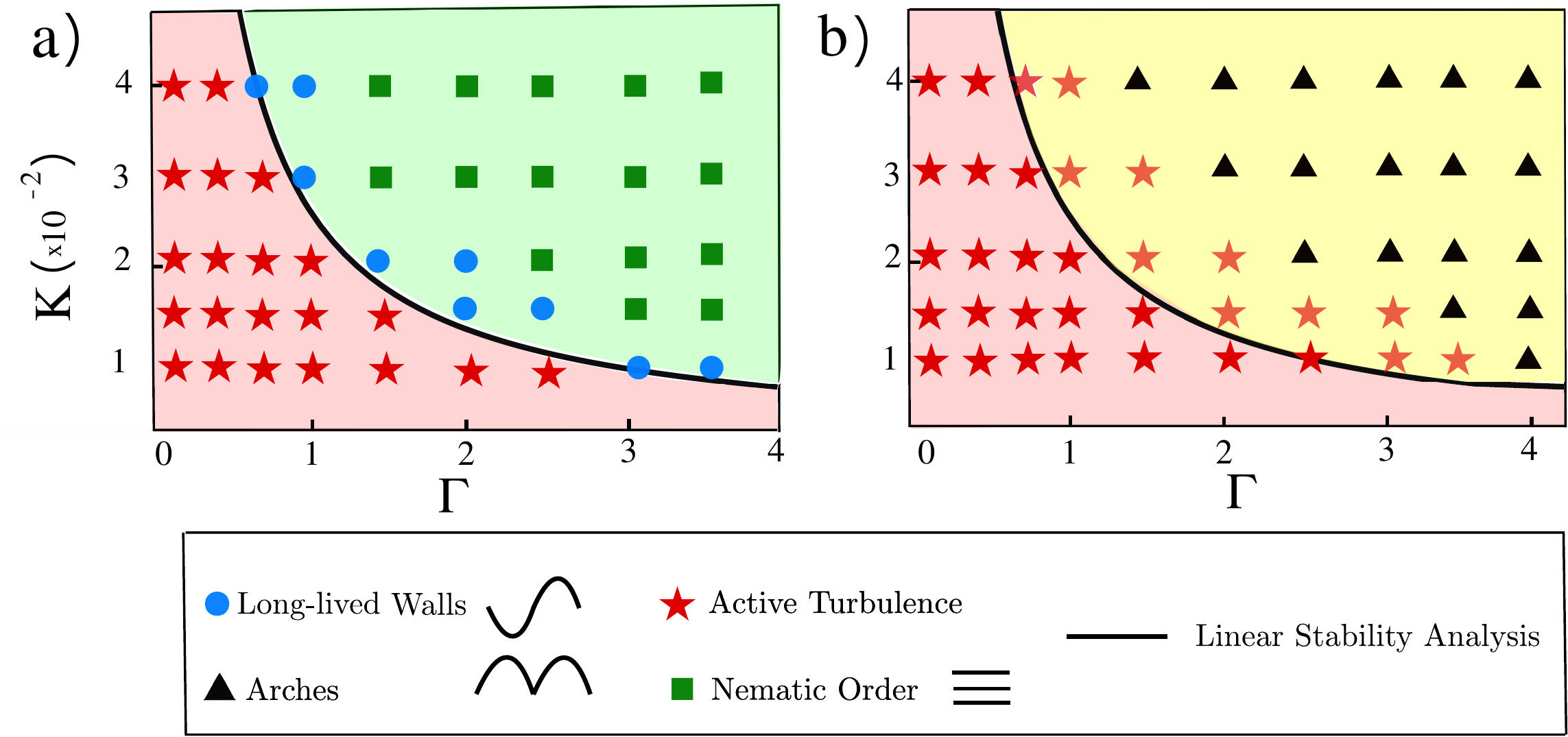}
  \caption{{\bf Stability diagrams as a function of elastic constant, $K$, and friction $\Gamma$:} (a) small noise in the initial conditions. Defects cannot form at high friction and the nematic state is stable. (b) large noise in the initial conditions. Defects that are present initially lead to arches which persist as a metastable state.
  The prediction of the linear stability analysis in the limit where the friction force dominates viscous forces, shown by black solid lines, agrees very well with the simulations. Note that when the initial noise is small, we observe long-lived walls for parameters close to the boundary of active turbulence (see Movie $10$). These structures are stable to defect formation during the simulations ($4 \times 10^6$ time-steps).}
  \label{archnew2}
\end{figure*}

Eq.~(\ref{sa}) shows that the longitudinal perturbations relax to zero and Eq.~(\ref{sa2}) gives the growth rate of transverse perturbations  $\tilde{Q}_{xy}$  as
\begin{eqnarray}
\omega =-q^2 \left\{S_0 \zeta \cos 2\theta (2\Gamma +2\eta q^2)^{-1}+ K \gamma \right\}.
\label{lstb}
\end{eqnarray}
For $\omega<0$, perturbations die out over time whereas for $\omega>0$ perturbations grow, the nematic state is unstable and active turbulence develops. 
Eq.~(\ref{lstb}) also confirms that the instability of the ordered state in extensile ($\zeta<0$) and contractile ($\zeta>0$) suspensions is caused by bend and splay perturbations, respectively. In the limiting case of zero friction, $\Gamma=0$, the growth rate simplifies to the well-known long wavelength hydrodynamic instability of active suspensions~\cite{ramaswamy2003active}.
The results of the linear stability analysis are consistent with the simulations of the full nonlinear equations
as shown in Fig.~\ref{archnew2}. 

In the phase diagram presented in Fig. \ref{archnew2} (b), we observe polar defect ordering in the arch phase, and the defect ordering that we observe leads to arches at very long time. Close to the boundary of the arch phase (intermediate value of friction), we observe polar defect ordering with larger number of defects and defect nucleation. \\

 Recent studies of active nematics have shown that the choice of flow-tumbling and flow-aligning regimes can have a profound effect on defect dynamics in active turbulence and in the presence of anisotropic friction~\cite{thijssen2020active,thijssen2020binding}. The measurements and analytical results of this paper are mainly for $\lambda=0$, since we expect the flow-tumbling regime to be more relevant deep in the ordered phase where $S \approx 1$. This is because the condition for being in the flow-tumbling regime ($ (3S+4) |\lambda| /(9 S) <1$) simplifies to $7 |\lambda|/9 <1$ deep in the ordered phase. This condition is always satisfied considering the fact that the tumbling parameter is in the interval $-1<\lambda<1$. To check this assumption, we examined the role of friction in systems with non-zero values of tumbling parameter $\lambda=\pm 0.7$. We found that for the negative (positive) value of the tumbling parameter, the transition to the arch-state in extensile systems in Fig. \ref{archnew2} (b) occurs at a smaller (larger) value of friction. To be more precise, using the same values of parameters as introduced at the beginning of section \ref{lowT}, the transition to arch patterns occurs at $\Gamma \sim 3.2$ for $\lambda=0$, and at $\Gamma \sim 4.1$ ($\Gamma \sim 2$) for $\lambda=0.7$ ($\lambda=-0.7$). When friction is small and arch patterns are not stable, for $\zeta \lambda>0$ two $+1/2$ defects come close to each other while pointing in the opposite direction similar to the recent study in \cite{thijssen2020binding}. However, when friction is large enough to stabilise arch patterns, we still observe polar order of $+1/2$ defects for all the choices of flow tumbling parameter. These observations support our choice of $\lambda=0$ in this paper.

\section{Arch Solutions}\label{metastability}

We next investigate the arch patterns that are observed at high friction and show that they correspond to metastable steady states of the equations of motion. We assume that the emerging structures are invariant in the $y$-direction, and that the friction force dominates the viscous force.
 
In the case of $\lambda=0$, Eq.~(\ref{eqn:ns}) can be solved for an incompressible system in the low-Reynolds number regime to give the pressure and velocity fields
\begin{align}
   &P = c_1 y -\frac{\zeta}{\Gamma} Q_{xx},\label{p}\\
   & u_y= \frac{\zeta}{\Gamma}\partial_x Q_{xy}+c_1, \:\:\: u_x=0. \label{uy}
\end{align} 
Since the system is invariant in the $y$ direction, $c_1=0$. Assuming that the magnitude of the nematic order is constant in space, we use Eq.~(\ref{uy}) together with Eq.~(\ref{n:qevoln}) to find the evolution of the orientation field:
\begin{align}
   &\partial_t \theta(x,t) = \frac{\zeta S_0}{4 \Gamma}\partial_x^2 \sin 2 \theta(x,t) + \gamma K \partial_x^2 \theta(x,t).\label{evut}
\end{align}
For zero activity ($\zeta=0$), the steady state solution of Eq.~(\ref{evut}) satisfies $\gamma K d_x^2 \theta(x)=0$.

Therefore, for periodic boundary conditions on a box of size $L$ so that $\textbf{Q}(x=0)=\textbf{Q}(x=L)$, 
 \begin{align}
   \theta(x)=\frac{m \pi}{L} x+\pi/2, \:\:\: m \in \text{integers}. \label{the}
\end{align} 
These solutions, which minimise  the elastic free energy, correspond to $m$ arches of equal width.
The global minimum is $m=0$, the nematic state, and arch solutions with $m\neq 0$, are local minima. As such, they  can be observed in systems with small noise but will be destabilised by larger values of the noise.

 \begin{figure*}
\centering
  \includegraphics[width=0.9\linewidth]{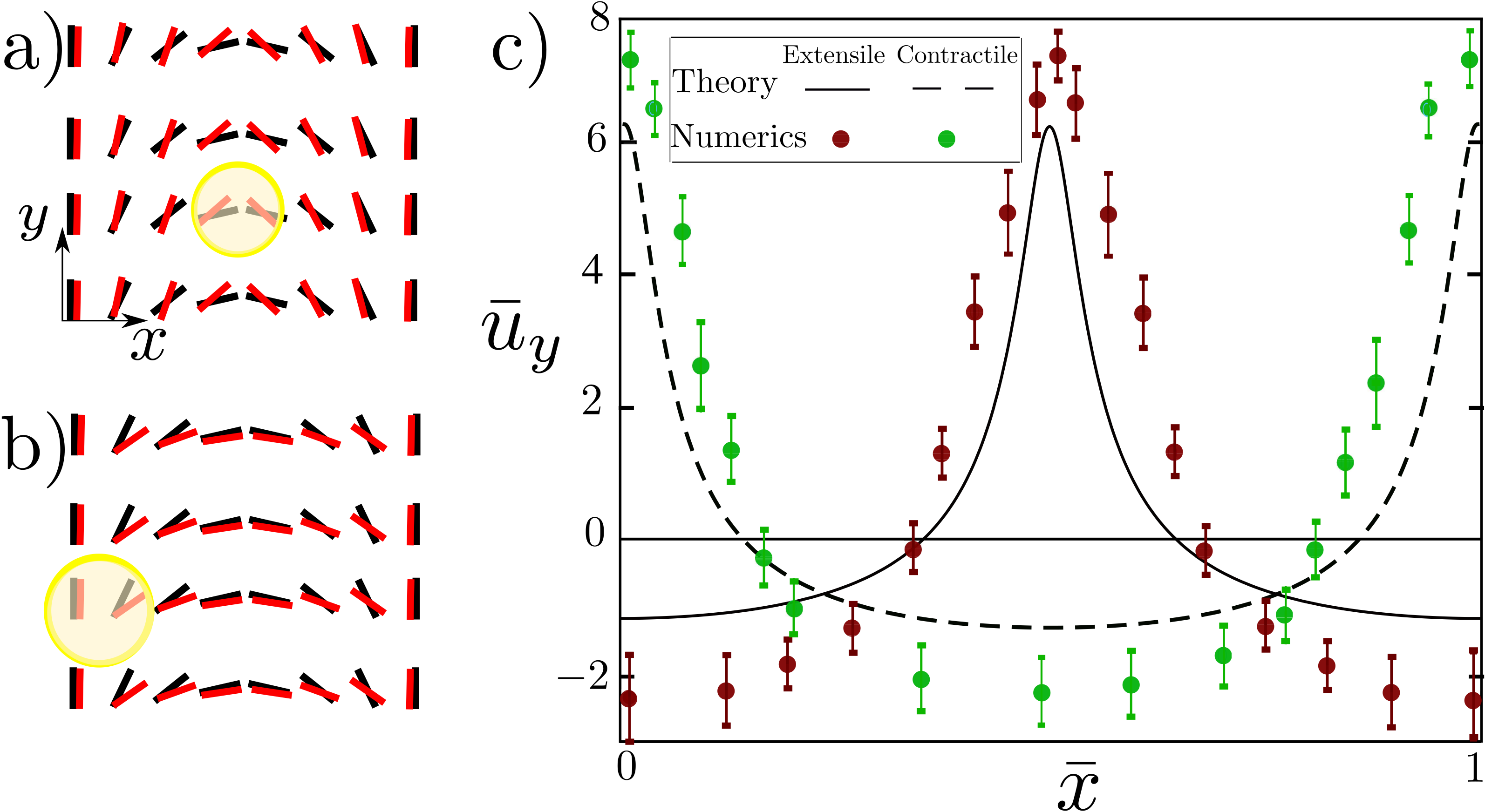}
  \caption{{\bf Arches:} Comparison of the director configuration of arches at zero activity (black) and finite activity (red) in (a) extensile and (b) contractile systems. 
  The yellow circles show that  extensile (contractile) activity increases bend (splay) deformations in arches. c) A comparison of the analytical solutions for the velocity field around arches (Eq.~(\ref{uy})) with the numerical solution of the full nematohydroynamic equations. The maximum velocity corresponds to the bend (splay) region of the arches in extensile (contractile) systems and points parallel to the polar axis of the arches in both cases.}
  \label{patterns2}
\end{figure*}
In the presence of activity, and assuming the same boundary conditions, the steady state director field satisfies
\begin{align}
   \frac{\zeta S_0}{2 K \gamma \Gamma} \sin 2 \theta(x) + \theta(x) = \frac{m \pi x}{L}+\frac{\pi}{2}. \label{steadystate3}
\end{align}

Eq.~(\ref{steadystate3}) does not have an analytic solution, but for $ |\zeta S_0| < 2 K \gamma \Gamma$ it has real solutions that 
can be found numerically. The solution of Eq.~(\ref{steadystate3}) is plotted in Fig. \ref{patterns2} (a) and (b) for extensile and contractile systems, respectively.
 
In particular, comparing to passive arches, in extensile (contractile) systems arches have larger bend (splay) deformations 
(see  Fig.~\ref{patterns2}). Note that the solutions of Eq.~(\ref{steadystate3}) correspond to $m$ arches with equal width which explains the observation in our simulations that, after creation, arches readjust towards equal widths (see Movie $9$).

Having in hand the director field, we can use Eq.~(\ref{uy}) to find the velocity field on arches. In Fig.~\ref{patterns2}(c) we compare the solution for $\bar{u}_y= u_y \ell^*/(\gamma K)$ from Eq.~(\ref{uy}), where $\ell^*$ is the arch width, with the flow field of arches obtained from the numerical solution of the full nematohydrodynamic equations, showing good agreement. Since the bend (splay) distortions are larger in arch solutions of extensile (contractile) systems, the maximum of the velocity field in extensile (contractile) systems is at the bend (splay) region of the arches. In addition, the maximum velocity points along the polar axis of arches in both contractile and extensile systems.

\section{Comparison to dry models }\label{comparison}

Intriguingly, arches, trails and polar ordering of defects reminiscent of the dynamics reported here, have been observed in dry systems. Here we compare our results to other work in the literature.\\

DeCamp {\it et al.}~\cite{decamp2015orientational} performed Brownian dynamics simulations of spherocylinders. The rods extended along their length and split, thus introducing extensile activity, and topological defects were created with their axes perpendicular to the local nematic field. As defects moved through the system they created long-lived trails  in the ordering of the spherocylinders which affected the motion of other defects leading to polar defect ordering.\\

Ref.~\cite{patelli2019understanding} includes a study of the microscopic dynamics of a Viscek-like model of self-propelling point particles that reverse their direction of motion with a given rate and interact through both aligning and repulsive interactions. The authors identify arch states which are formed by the motion of topological defects and which coexist with the nematic state.  The arches do not have a preferred size, but evolve towards equal widths as observed here. The arches lead to the polar ordering of $+1/2$ defects close to the border of the arch region where it is still possible to create defect pairs.
Patelli~{\it et al.}~\cite{patelli2019understanding} also address the continuum limit of their model. They find arches to be metastable over a larger region of phase space, and argue that this is because these are {\it fragile} states, easily destroyed by the noise in the microscopic simulations. No defect ordering is seen in the continuum model. An important conclusion of 
Ref.~\cite{patelli2019understanding} is to point out the sensitivity of defect properties to parameter details.\\

Several authors~\cite{srivastava2016negative,putzig2016instabilities,oza2016antipolar} have considered 
 a dry system as a limiting case  of the continuum equations considered here, where friction completely dominates viscous effects. Allowing the system to be compressible, it is then possible to slave the velocity field to the active force and to describe the dynamics in terms of a single equation for the nematic tensor $\bf Q$. Srivastava {\it et al.}~\cite{srivastava2016negative} showed that the equation of motion contains a term, of the same form as the elasticity, which can drive the effective elastic constant negative. They interpret the appearance of arch states (which they term kink walls) as a consequence of the negative effective elastic constant, and predict a transition between both the nematic and isotropic
 states and an arch state by increasing activity. Here we find that arches co-exist only with the nematic phase. Srivastava {\it et al.} see defect ordering near the boundaries of the arch phase. 
 Putzig {\it et al.}~\cite{putzig2016instabilities} extend the arguments to a system that breaks Galilean invariance by weighting the convective and rotational terms in the equation of motion differently.
 They find polar ordering of topological defects if Galilean invariance is broken but the ordering ceases to exist in the Galilean invariant case. The authors argue that $+1/2$ defects form trails in the form of arches. Other $+1/2$ defects tend to reorient and follow the trails, leading to polar ordering. Similarly to our work, the arch patterns coexist  with the nematic phase and are not caused by an instability in the homogeneous nematic state. However, in their system, arch structures appear in a small region of the phase diagram, whereas in our model they cover a large region of parameter space.
 Similar equations are considered in Oza and Dunkel~\cite{oza2016antipolar}  with the difference that they ignore any rotational effect of the flow. Their elastic constant is negative by definition, and they stabilise the system by introducing higher order derivatives into the free energy, thus introducing a length scale.  They find several different states including  defect-free ground-states,  long-lived, lattice-like configurations of defects with anti-polar order at low activities and arches. \\

Ref.~\cite{shankar2019hydrodynamics} provides an analytic approach to formulating the coarse-grained equations of motion of a compressible, charge-neutral system of $\pm1/2$ defects in the over-damped regime where friction is dominant over viscous dissipation, and in the presence of noise. This formalism predicts active turbulence, together with a continuous transition to polar defect ordering at higher activities and for slow relaxation of the nematic field. The ordering is explained as being due to the torques exerted on defects as they interact with the kink-wall (arch) director field configurations left by unbinding $\pm 1/2$ pairs. The arch patterns are only observed when the noise is sufficiently large that defects can nucleate and move in the system. This is in agreement with our work where we only find arches when noise is large in our initial condition. However, we find arches as a steady-state solution only when activity is smaller than a critical value ($|\zeta|<|\zeta_c|= 2 \gamma K \Gamma$), 
whereas in ~\cite{srivastava2016negative} the transition to polar defect ordering and arches is predicted to occur at large activities.

\section{Summary and Conclusion}\label{discussion}

We have studied the role of friction in the nematic phase of active incompressible suspensions. Friction introduces a memory into the system such that the motion of topological defects leaves trails, arch-like distortions in the director field that persist for a time that increases with increasing friction. Defects in both extensile and contractile systems move in the same direction with respect to the polar axis of arches. They  interact with the trails left by other defects, leading to the polar order of $+1/2$ defects observed in microscopic simulations \cite{decamp2015orientational}.
Moreover, the polar nature of arches can break the symmetry of $-1/2$ defects, allowing them to self-propel along these polar patterns.

At very high friction there is insufficient energy to create new topological defects but any defects already in the system create arches in the director field
 before eventually annihilating. In the absence of defects the arches align parallel and readjust to equal widths.
We have shown, analytically and numerically, that regular arch patterns are steady-state solutions of the nematohydrodynamic equations at high friction which coexist with the nematic phase \cite{patelli2019understanding}. 
In the simulations, for small values of the initial noise, there are no defects and the nematic state is stable.  However, when the initial noise is large, the initial configuration contains defects which nucleate the arch state.

We have also presented a summary of the similarities and differences in the behaviour of systems described by active nematohydrodynamics with friction to previous results for related models. In particular our study shows that compressibility is not necessary for the formation of arches but that arches also exist in the high friction regime of wet, incompressible systems.

For completeness we mention how friction changes the behaviour of an active nematic where the passive limit corresponds to an {\it isotropic} phase. Here the local nematic order needed for active turbulence is induced by the active stresses themselves. Hence the nematic order parameter is small, elastic energies are small, and it is easier to create topological defects even at high friction. As friction is increased and the active flow is screened these tend to order into regular defect lattices \cite{lattices}. 

Our work could motivate new studies on whether memory effects could be used to control the behaviour of topological defects in active nematics. 
It was shown in Ref. \cite{maitra2018nonequilibrium} that an active system in contact with a substrate can support an additional active force that stabilises nematic order and it would be interesting to investigate this further.
It is also interesting to speculate whether memory effects could have relevance in biological systems, such as bacterial colonies or epithelial cell sheets, which can be modelled as active nematics with friction.

\section*{Conflicts of interest}
There are no conflicts to declare.

\section*{Acknowledgements}
M. R. N. acknowledges the support of the Clarendon Fund Scholarship. A. D. acknowledges support from the Novo Nordisk Foundation (grant No. NNF18SA0035142), Villum Fonden (Grant no. 29476), Danish Council for Independent Research, Natural Sciences (DFF-117155-1001), and funding from the European Union’s Horizon 2020 research and innovation program under the Marie Sklodowska-Curie grant agreement No. 847523 (INTERACTIONS).

\balance


\bibliography{rsc} 
\bibliographystyle{rsc} 

\end{document}